\documentclass[aps,prb,twocolumn,superscriptaddress]{revtex4}
\usepackage{textcomp}
\usepackage{amsmath}
\usepackage{amssymb}
\usepackage{graphicx,graphics}
\usepackage{latexsym,verbatim}
\usepackage{dcolumn} 
\usepackage{color}
\usepackage{cancel}
\usepackage{ulem}
\usepackage{url}

\begin{document}

\title{Surface chiral superconductivity in odd-parity nematic superconductors with magnetic impurities}

\author{Luca Chirolli}
\email{l.chirolli@berkeley.edu}
\affiliation{Department of Physics, University of California, Berkeley, CA-94720}
\affiliation{Instituto Nanoscienze-CNR, I-56127 Pisa}

\begin{abstract}
We study odd-parity nematic superconductivity in doped topological insulators in presence of surface magnetic impurities. The peculiar surface subgap spectrum, characterized by a Majorana flatband, nodal cones and the surface states of the parent topological insulator, gives rise to an overall ferromagnetic RKKY interactions between the surface impurities. An additional coupling between the impurities and a preemptive chiral order parameter promote a surface time-reversal symmetry breaking solution at the surface of the system. We discuss the relevant scenarios and suggest to engineer surface chiral superconductivity by properly choosing magnetic adatoms with highly anisotropic exchange coupling. 
\end{abstract}

\maketitle

\section{Introduction}
 
Chiral superconductivity is a highly interesting and long sought unconventional state of matter that spontaneously breaks time-reversal symmetry through the development of a Cooper pair finite angular momentum \cite{sigrist1991phenomenological,kallin2016chiral}. It represents an instance of topological superconductivity \cite{qi2011topological,ando2015topological,sato2017topological}, that has attracted great interest thanks to its potential for hosting Majorana fermions in vortex cores \cite{alicea2012new,beenakker2013search,elliot2015colloquium}, and in topological quantum computation \cite{nayak2008non-abelian,dassarma2015majorana,chirolli2018chiralmajorana}. Intrinsic chiral superconductivity is an unstable state of matter and its occurrence has been suggested in particular conditions, such as layered material like UPt$_3$ \cite{tou1998nonunitary}, Li$_2$Pt$_3$B \cite{nishiyama2007spin}, Sr$_2$RuO$_4$ \cite{maeno1994superconductivity,mackenzie2017even}, SrPtAs \cite{biswas2013evidence} and 4Hb-TaS$_2$ \cite{ribak2020chiral}. However, its detection relies on observation of spontaneous magnetization or generation of local magnetic fields \cite{kvorning2018proposed}, that is usually hindered by Meissner screening, and its unequivocal demonstration still remains controversial.

Quantum design has become a very attractive and promising way to attain unconventional and fascinating states of matter. This is the case of engineered topological superconductors \cite{qi2011topological,alicea2012new,beenakker2013search}, where by bringing together materials with different properties it is possible to engineer the resulting compound at will.  It is then natural to wonder whether chiral superconductivity can be stabilized by suitable quantum design. To this end the relevant ingredients that need to be brought together are the quasi two-dimensional character, a time-reversal symmetry breaking (TRSB) phase trigger, and a multi-component order parameter (OP) \cite{sigrist1991phenomenological}. A bulk two-component OP can choose two solutions, either a rotation symmetry breaking solution, the nematic state, or a chiral TRSB solution. In three-dimensional Dirac materials with closed weakly anisotropic Fermi surface the nematic solution is more stable \cite{fernandes2012preemptive,fu2014odd-parity,chirolli2017time-reversal,scheurer2017selection}.  Nevertheless, $C_3$ crystal symmetry \cite{scheurer2017selection} and two-dimensionality \cite{chirolli2018chiral,yang2019the} help in stabilizing a chiral solution, and magnetic fluctuations \cite{chirolli2017time-reversal,yuan2017superconductivity-induced} can provide a mechanism that triggers a TRSB phase. However, none of them alone is sufficient nor fully practical.

\begin{figure}[b]
\includegraphics[width=0.45\textwidth]{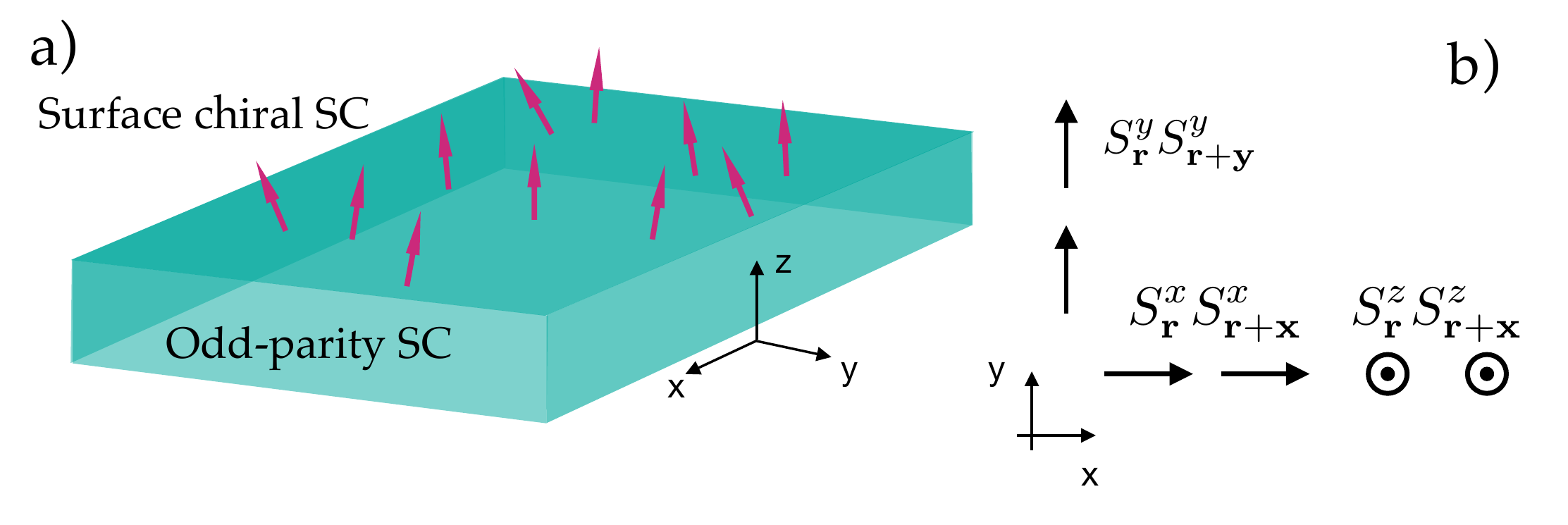}
\caption{a) Schematics of the setup considered: a bulk odd-parity nematic superconductor with surface magnetic disorder and chiral surface solution. b) Directional dependence of the RKKY interaction experienced by magnetic impurities.  
\label{Fig1}}
\end{figure}

In this work, we consider an odd-parity nematic superconductor in presence of surface magnetic impurities. The system is schematized in Fig.~\ref{Fig1}a). We study the Ruderman-Kittel-Kasuya-Yosida (RKKY) interaction mediated by the surface gapless states. Three main actors contribute to the interaction: i) states close to the nodes of the gap, ii) a flat band of Majorana surface states extending between the nodes, and iii) the surface states of the parent TI.  We find that for an impurity ensemble dilute on the scale of the Fermi wavelength a ferromagnetic interaction is mediated by the surface gapless modes and the system, albeit disordered, is expected to show ferromagnetic order. Close to the surface, a preemptive chiral OP couples to the out-of-plane magnetization. For an in-plane magnetic order, fluctuations of the out-of-plane magnetizations generated by the chiral OP itself promote a phase transition to a TRSB surface state for sufficiently strong coupling. For an out-of-plane order, the chiral OP always condenses at the surface. Due to the small scales provided by the SC gap and in the dilute impurity ensemble approximation, the RKKY mediated in-plane order scenario turns out to be quite fragile and in general the out-of-plane order is realized. These results open the way to engineering surface chiral superconductivity in bulk nematic odd-parity superconductors and provide a mechanism to stabilize the chiral phase in thin samples. 

A promising platform for the realization of surface chiral superconductivity is provided by doped Bi$_2$Se$_3$ \cite{chirolli2018chiral,yang2019the}. Early experiments \cite{sasaki2011topological,hor2010superconductivity,sasaki2014superconductor,liu2015superconductivity} and recent measurements  \cite{matano2016spin-rotation,asaba2017rotational,yonezawa2016thermodynamic,nikitin2016high-pressure,willa2018nanocalorimetric,pan2016rotational,shen2017nematic,andersen2018nematic,smylie2018superconducting,kawai2020direction} have by now established the odd-parity nematic character of the superconducting state, characterized by a $C_2$ symmetry. The latter  is consistent with the two-component $E_u$ representation of the $D_{3d}$ crystal point group of the material \cite{FuBerg,fu2014odd-parity,venderbos2016odd-parity,venderbos2016identification}, possibly triggered by odd parity fluctuations \cite{wu2017nematic,hecker2018vestigial}, density wave fluctuations \cite{kozii2019nematic}, structural distortion \cite{kuntsevich2018structural}, nematicity above $T_c$ \cite{sun2019quasiparticle}, and ferroelectric fluctuations \cite{kozii2019superconductivity}. 

The results presented are generic of odd-parity nematic superconductors, and can be extended to other systems such as   UPt$_3$ \cite{sauls1994theorderparameter,joynt2002thesuperconducting}, Sr$_2$RuO$_4$ \cite{rice1995sr2ruo4,huang2018possible,mackenzie2003thesuperconductivity} or topological semimetals \cite{siddiquee2019nematic}, rendering these systems an ideal platform for quantum designing of unconventional physics.

\section{The system}

We start the analysis considering the $k\cdot p$ unperturbed Hamiltonian describing doped Bi$_2$Se$_3$. The latter is well described by a 3D anisotropic massive Dirac equation ($\hbar=1$) \cite{FuBerg}
\begin{equation}\label{Eq:H0}
{\cal H}^0_{\bf k}=m\sigma_x+v(k_xs_y-k_ys_x)\sigma_z+v_zk_z\sigma_y,
\end{equation}
where the Pauli matrices $\sigma_i$ span a two-fold orbital subspace and $s_i$ are spin Pauli matrices. Superconductivity is studied by means of the Bogolyubov-deGennes (BdG) Hamiltonian  $H=\frac{1}{2}\sum_{\bf k}\psi^\dag_{\bf k}{\cal H}_{\bf k}\psi_{\bf k}$. In the Nambu basis $\psi_{\bf k}=({\bf c}_{\bf k},is_y{\bf c}^\dag_{-{\bf k}})^T$, with ${\bf c}_{\bf k}$ a vector of fermion operators in spin and orbital basis, the BdG Hamiltonian reads
\begin{equation}\label{HBdG-oddparity}
{\cal H}_{\bf k}=({\cal H}^0_{\bf k}-\mu)\tau_z+\hat{\Delta}\tau_++\hat{\Delta}^\dag\tau_-.
\end{equation} 
In the $E_u$ odd parity channel the gap matrix reads $\hat{\Delta}=-\psi_x\sigma_ys_y+\psi_y\sigma_y s_x$, where $\boldsymbol{\psi}=(\psi_x,\psi_y)$ is the two-component OP.  The ground state admits two possible solutions: i) a TR invariant nodal nematic phase $\boldsymbol{\psi}\propto (1,0)$ and ii) a chiral phase $\boldsymbol{\psi}\propto (1,\pm i)$, that breaks TR symmetry. The chiral phase has Weyl nodes in 3D and is fully gapped in 2D systems. Consistently with experiments, we choose a bulk nematic phase. 

\begin{figure}[t]
\includegraphics[width=0.45\textwidth]{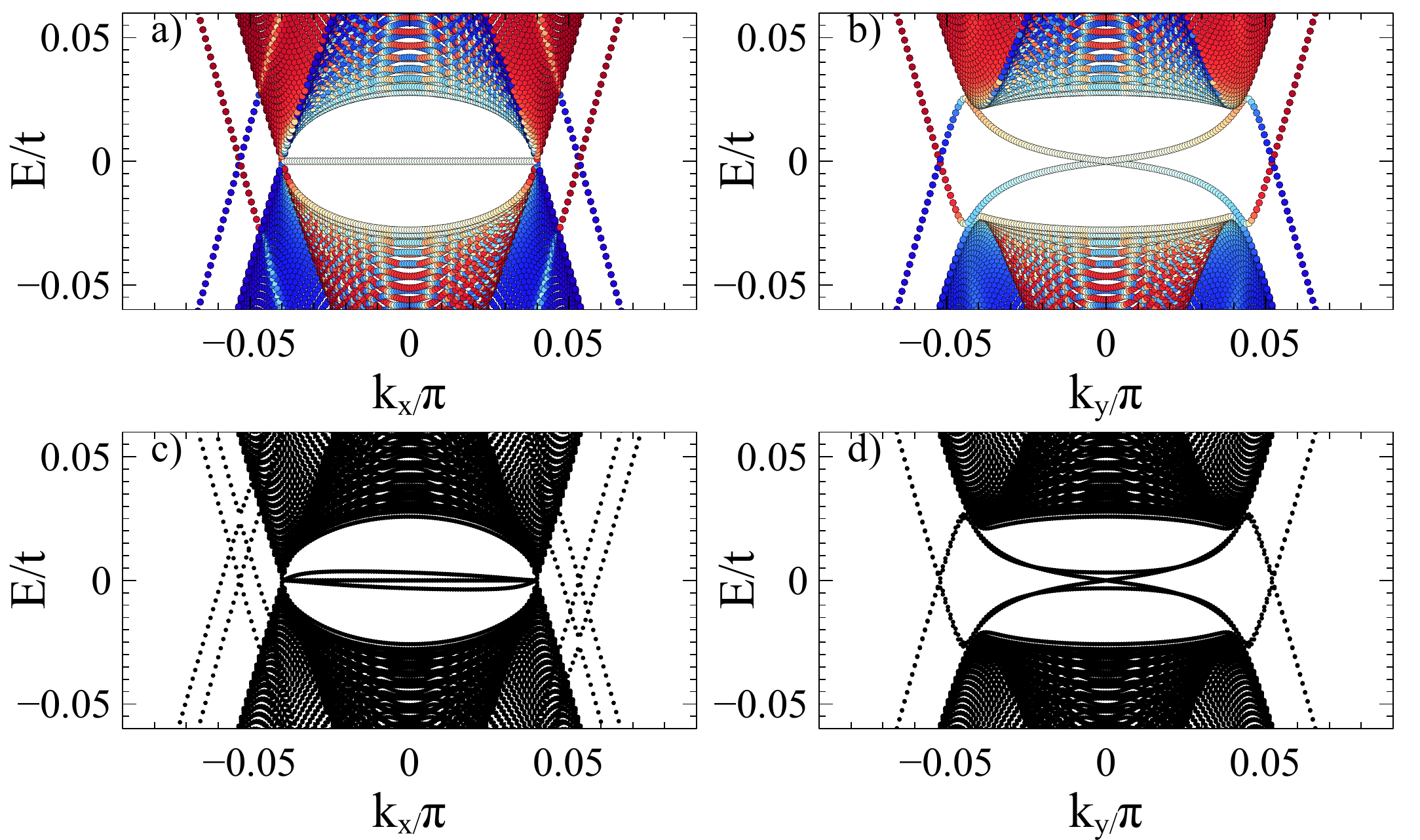}
\caption{Surface spectrum of the system obtained with a tight-binding model for a slab of 200 bilayers \cite{Note1}. a) and b) in the nematic phase along the $k_x$ and $k_y$, respectively. The color code represents the charge (red electrons, blue holes). c) and d) in the nematic phase with surface chiral solution and finite magnetization $\langle S_y\rangle$ only on one surface, along the $k_x$ and $k_y$, respectively. 
\label{Fig2}}
\end{figure}

We assume the system to occupy the $z>0$ region of space. The full surface spectrum obtained by a tight-binding model \footnote{See Supplemental Material for details about the tight-binding model, the RKKY interaction from the nodes, and the Ginzubrg Landau free energy.} is shown in Fig.~\ref{Fig2}a) and \ref{Fig2}b) and nodes are present in the spectrum. At the surface of the system, a topologically protected, doubly degenerate Majorana flat band appears for $|k_x|<k_F=\sqrt{\mu^2-m^2}/v$, extending between the surface projection of the bulk nodes at $\pm k_F$. Additional crossing takes places at momentum $\pm \mu/v$. These states are gapless modes originating from the TI surface states, that cross the Fermi level at finite momentum and are hybridized but not gapped by the odd-parity OP \cite{hsieh2012majorana}.

We then place magnetic impurities on the $z=0$ surface of the system and assume coupling to the electrons via an anisotropic exchange interaction 
\begin{equation}\label{Exchange}
{\cal H}_Z=-\frac{1}{n}\sum_i\left[J_zS^z_is_z+J_\parallel(S_i^xs_x+S^y_is_y)\right],
\end{equation}
where ${\bf S}_i$ is the spin of the impurity located at position ${\bf r}_i$, ${\bf s}$ is the electronic spin operator, $J_z$ and $J_\parallel$ are out-of-plane and in-plane exchange couplings, and $n=N/V$ is the electron density \cite{abrikosov-fundamentals}. Impurities also induce scattering via the scalar part of their potential. This typically has detrimental effects of unconventional pairing due to momentum randomization. Nevertheless, for sufficiently diluted impurities, such that the mean free path $\ell_{\rm mf}$ is much larger than the Fermi wavelength $\lambda_F$ but comparable to the coherence length $\xi$, $\lambda_F\ll \ell_{\rm mf}\sim \xi$, we neglect their effect.

\section{RKKY interaction} 

For relatively weak exchange coupling, we integrate away the fermionic degrees of freedom and obtain the RKKY interaction experienced by the magnetic impurities \cite{abrikosov-fundamentals},
 \begin{equation}
\chi_{\mu\nu}({\bf r})=\frac{J_\mu J_\nu}{n^2} T\sum_{i\omega_n}{\rm Tr}\left[s_\mu G_{i\omega_n}({\bf r})s_\nu G_{i\omega_n}(-{\bf r})\right],
\end{equation}
where $G_{i\omega_n}({\bf r})=\sum_{\bf k}e^{i{\bf k}\cdot{\bf r}}(i\omega_n-{\cal H}_{\bf k})^{-1}$ is the Green's function of the BdG Hamiltonian Eq.~(\ref{HBdG-oddparity}).  Three main actors mediate the interaction: i) the states around the nodes, ii) the Majorana flat band, and iii) the surface hybridized TI modes. The TI states contribution can be estimated by neglecting the hybridization induced by the gap. In this case, well know results for doped TI surface states apply \cite{liu2009magnetic,biswas2010impurity-induced,garate2010magnetoelectric,zhu2011electrically,abanin2011ordering,zyuzin2014rkky,efimkin2014self-consistent}. The different terms that arise show oscillations at $2k_F$ and decay as $1/r^2$. In addition, above critical temperature, conduction band electrons provide an additional term that oscillates at $2k_F$ and decays as $1/r^3$. We neglect fast decaying terms in a diluite impurity ensemble approximation.

\subsection{RKKY Majorana flatband}
  
The effective Hamiltonian describing the Majorana flat band is written as 
\begin{equation}\label{Eq:HsurfFB}
h_{\bf k}=-v_Mk_y\hat{\alpha}_y,
\end{equation}
with $v_M=v m\Delta/\mu^2$ the velocity of the Majorana modes \cite{hsieh2012majorana} and $\hat{\alpha}_i$ a set of Pauli matrices spanning the subspace defined by $|\phi_\pm\rangle=\frac{1}{2}\left(\begin{array}{c}1\\ \pm i \end{array}\right)_s\otimes \left(\begin{array}{c}1\\ \mp i\end{array}\right)_\tau$. The RKKY interaction mediated by Majorana fermions has been discussed in Ref.~[\onlinecite{choy2013magnetic}] and it represents a particular case of surface TI fermions,  with two peculiar differences: i) the zero chemical potential condition is satisfied exactly and ii) only one spin component, the $s_y$ in this case, has nonzero projection on the Majorana wavefunction $|\phi_\pm\rangle$, resulting in the Pauli matrix $\hat{\alpha}_z$. This is the well known Ising property of Majorana Kramers pairs \cite{liu2009magnetic}. The flatband-mediated RKKY interaction reads  \cite{Note1}
\begin{equation}\label{Eq:RKKY-fb}
\chi^{\rm FB}_{yy}(x,y)=-\frac{J_\parallel^2\rho_{2D}}{2\pi v_M n^2}\frac{\sin^2(k_Fx)}{(k_Fx)^2}\frac{f(k_Fy)}{y},
\end{equation}
with $f(x)=(2/\pi)\int_0^xdz_1dz_2\cos(z_1-z_2)/(z_1+z_2)$ and $\rho_{2D}=k_F^2/\pi^2$ the surface density. The flatband extending between momenta $\pm k_F$ along $x$ generates a contact interaction that dies on a scale $1/k_F$. Along the $y$ direction, $f\to 1$ for large argument, so that for an ensemble  diluite on the scale of the Fermi wavelength $\lambda_F=1/k_F$, the interaction has a purely ferromagnetic long range character. As for the case of TI, magnetic order along the direction dictated by the relevant Majorana operator opens a gap in the Majorana spectrum (Fig.~\ref{Fig2}c) and d)). The RKKY interaction survives also in presence of a gap, self-consistently sustained by the interaction itself \cite{abanin2011ordering,efimkin2014self-consistent}. Additionally, a magnetization along $y$ acts as a tilting field \cite{chirolli2018magnetic,chirolli2018signatures} on the TI surface modes along $k_x$. 

\subsection{RKKY nodes} 
 
We then consider the RKKY interaction generated by the nodes at the surface. The BdG Hamiltonian Eq.~(\ref{HBdG-oddparity}), projected onto the conduction band states  $\{|\psi_{\rm cb}^1\rangle_{{\bf k}}, |\psi_{\rm cb}^2\rangle_{{\bf k}}\}$ and expanded around the nodal points at $\pm k_F$, reads \cite{Note1}
\begin{equation}
{\cal H}^\pm=\left(\begin{array}{cc}
\pm v_x k_x & \delta (k_y\tilde{s}_z-k_z\tilde{s}_y)\\
\delta (k_y\tilde{s}_z-k_z\tilde{s}_y) & \mp v_xk_x
\end{array}\right),
\end{equation}
with $v_x=v^2k_F/\mu$, $\delta=\Delta v/\mu$, and $\tilde{s}_i$ are Pauli matrices spanning the conduction band states. Introducing the rescaled position  $\boldsymbol{\rho}=\mu(x/(k_F\xi),y,zv/v_z)/v$, the resulting spin susceptibilities are given by
\begin{eqnarray}
\chi_{xx}(\rho)&=&-\chi_0J_\parallel^2\left[(m/\mu)^2A_0(\rho)+A_x(\rho)\cos(2k_Fx)\right],\nonumber\\
\chi_{yy}(\rho)&=&-\chi_0J_\parallel^2(1-(m/\mu)^2)A_z(\rho)\cos(2k_Fx),\nonumber\\
\chi_{zz}(\rho)&=&-\chi_0J_z^2(1-(m/\mu)^2)A_y(\rho),\nonumber
\end{eqnarray}
where $\chi_0=8\nu_F^2\Delta(\mu/k_Fv)^4/n^2$ and $\nu_F=\mu k_F/(2\pi^2 v v_z)$ is the density of states at the Fermi level of the bulk Hamiltonian. The functions $A_i(\rho)$ \cite{Note1} carry a weak dependence on the direction $\hat{\bf r}$ and are well approximated by $A_i(\rho)=\sin^3(\rho/2)/(3\rho^3)$.  The terms oscillating with frequency $2k_F$ originate from internode scattering, whereas the others come from intranode scattering. Along the nodal direction $x$, the length scale is provided by the superconducting coherence length $\xi=v/\Delta$, whereas along the other directions it is given by $\lambda_F$. Assuming a bulk gap $\Delta\sim 1~{\rm K}$ and a velocity $v=0.6\times 10^8~{\rm cm/s}$, we have $\xi\sim 5~\mu{\rm m}$. In turn, assuming $\mu=0.33~{\rm eV}$ and $m=0.3~{\rm eV}$, we have $\lambda_F\sim 3~{\rm nm}$. This way, for an impurity ensemble diluite on the scale of $\lambda_F$, the RKKY interaction mediated by the nodes acts only along the $x$ direction and its character is mainly ferromagnetic. 

\section{Impurity-Chiral order parameter coupling}

As shown in Refs.~[\onlinecite{chirolli2017time-reversal,yuan2017superconductivity-induced}], magnetic impurities couple to the chiral OP $i\boldsymbol{\psi}\times\boldsymbol{\psi}^*$, that transforms as a pseudovector and can be regarded as an electron spin polarization \cite{zyuzin2017nematic} or Cooper pair spin. Although in the nematic state the chiral OP is zero, a coupling to magnetic impurity can trigger a finite value in proximity of the surface. Including the RKKY interaction arising from the Majorana flat band and the nodes in a total susceptibility $\chi^{\mu\mu}$, the free energy describing magnetic impurities coupled to the order parameter reads
\begin{equation}
F_{m}=\sum_{ij}\chi^{\mu\mu}(i,j)S^\mu_iS^\mu_j+i\frac{J_z\kappa}{n}\sum_iS^z_i(\boldsymbol{\psi}_{0,i}\times\boldsymbol{\psi}_{0,i}^*)_z,
\end{equation}
where $\kappa\simeq \mu \nu_F/T_c^2$ is calculated in the normal state [\onlinecite{chirolli2017time-reversal}]. 

The magnetic interaction is of XX and ZZ type along the $x$ direction and of YY type along the $y$ direction (see Fig.~\ref{Fig1}) and the chiral OP plays the role of an external field pointing about the $\hat{z}$ direction. Whereas the oscillations with frequency $2k_F$ tend to randomize the XX and YY coupling, the ZZ coupling is practically constant for $x<\xi$. The impurity ensemble is in general disordered, so that the values of $\chi^{\mu\mu}(i,j)$ can be thought as random in magnitude, distributed about different non-zero negative average values. The system belongs to the widely studied class of spin glass models with ferromagnetic random couplings  \cite{binder1986spin,mezard1987spin}. The ground state is ferromagnetic \cite{gabay1981coexistence}, with the total magnetization pointing about the direction of largest average coupling $\chi^{\mu\mu}$.  We study two cases: i) non negligible interaction with in-plane order, for $\ell_{\rm mf}\lesssim\xi$ and ii) negligible interaction  for $\ell_{\rm mf}\gtrsim \xi$ or out-of-plane order. 

For $J_\parallel> J_z$, in-plane ferromagnetic order is expected. The preemptive chiral OP tends to destroy the in-plane order and establish a non zero expectation value of $\langle S^z_i\rangle$, proportional to the chiral OP itself, $\langle S^z_i\rangle=(\Theta \kappa J_z/n)|\boldsymbol{\psi}_{0}\times\boldsymbol{\psi}_{0}^*|$, where $\Theta$ is the zero field susceptibility in the ferromagnetic phase. This yields a second order correction  to the superconductor free energy
\begin{equation}\label{Eq:IntJparallel}
F=F_\psi-\Theta(\kappa J_z/n)^2n_{\rm imp}|\boldsymbol{\psi}_0\times\boldsymbol{\psi}_0^*|^2,
\end{equation} 
with $n_{\rm imp}$ the impurity concentration. For a spin chain with nearest neighbor coupling $\chi_0 J_\parallel^2$, the susceptibility is $\Theta=1/(2\chi_0J^2_\parallel)$, and the  interaction is $\propto (J_z/J_\parallel)^2$.

For $J_z> J_\parallel$ the ferromagnetic order is out-of-plane, the ground state has already a finite $\langle S^z_i\rangle$ and the correction to the free energy is linear in the chiral OP,
\begin{equation}\label{Eq:IntZ}
F=F_\psi-i\kappa J_z(n_{\rm imp}/n)\boldsymbol{\psi}_0\times\boldsymbol{\psi}_0^*,
\end{equation} 
where we assumed the ground state with all impurities pointing about the $\hat{z}$ direction. This scenario also applies to the experimentally relevant case in which the impurity can be considered as non interacting.

\section{Surface chiral solution}

In a semi-infinite system it is natural to expect a TRSB solution in proximity of the surface, so that $\boldsymbol{\psi}$ acquires a position dependence that matches two asymptotic solutions, a nematic one at infinity and a TRSB one at $z=0$. We describe the modulation of the OP via a Ginzburg-Landau (GL) free energy whose form is dictated by symmetry arguments,
\begin{equation}\label{Eq:GL}
F_\psi=\int \frac{d^3{\bf r}}{V} \left[a|\boldsymbol{\psi}|^2+b|\boldsymbol{\psi}|^4+b'|\boldsymbol{\psi}\times\boldsymbol{\psi}^*|^2+
\beta_z|\partial_z\boldsymbol{\psi}|^2\right],
\end{equation}
where $V$ is the volume of the system and we neglect in-plane gradients \cite{zyuzin2017nematic} . Below $T_c$, $a$ becomes negative and a finite $b>0$ ensure a stable finite solution.  The two possible nematic and chiral solutions are favoured by $b'>0$ and $b'<0$, respectively. In absence of TRSB perturbations, the condition $b'>0$ is met for bulk 3D systems.

We then parametrize $\boldsymbol{\psi}$ in terms of real valued amplitude $\psi(z)$ and relative phase $\varphi(z)$, $\boldsymbol{\psi}=\psi(e^{-i\varphi/2},e^{i\varphi/2})/\sqrt{2}$ \footnote{This parametrization  identically satisfies $j_z=-2ei\beta_z\sum_{\mu=x,y}(\psi_\mu^*\partial_z\psi_\mu-(\partial_z\psi_\mu^*)\psi_\mu)=0$ and  guarantees zero current orthogonal to the surface \cite{zyuzin2017nematic}}. We rescale the amplitude by the bulk value $\psi_\infty\equiv \sqrt{|a|/(2b)}$, the position by the GL coherence length $\xi=\sqrt{\beta_z/(2|a|)}$, and set $\eta=b'/b$. For $\eta\ll 1$ we assume constant amplitude and the GL free energy is written as \cite{Note1}
\begin{equation}\label{Eq:deltaF}
\delta F\propto\int_{0}^\infty dx \left[{\cal F}(\varphi,\varphi')-g U(\varphi)\delta(x)\right],
\end{equation}
where ${\cal F}=(\varphi')^2/4+\eta U(\varphi)$. The potential $U$ depends on the boundary interaction.

\subsection{$J_\parallel>J_z$}

\begin{figure}[t]
\includegraphics[width=0.45\textwidth]{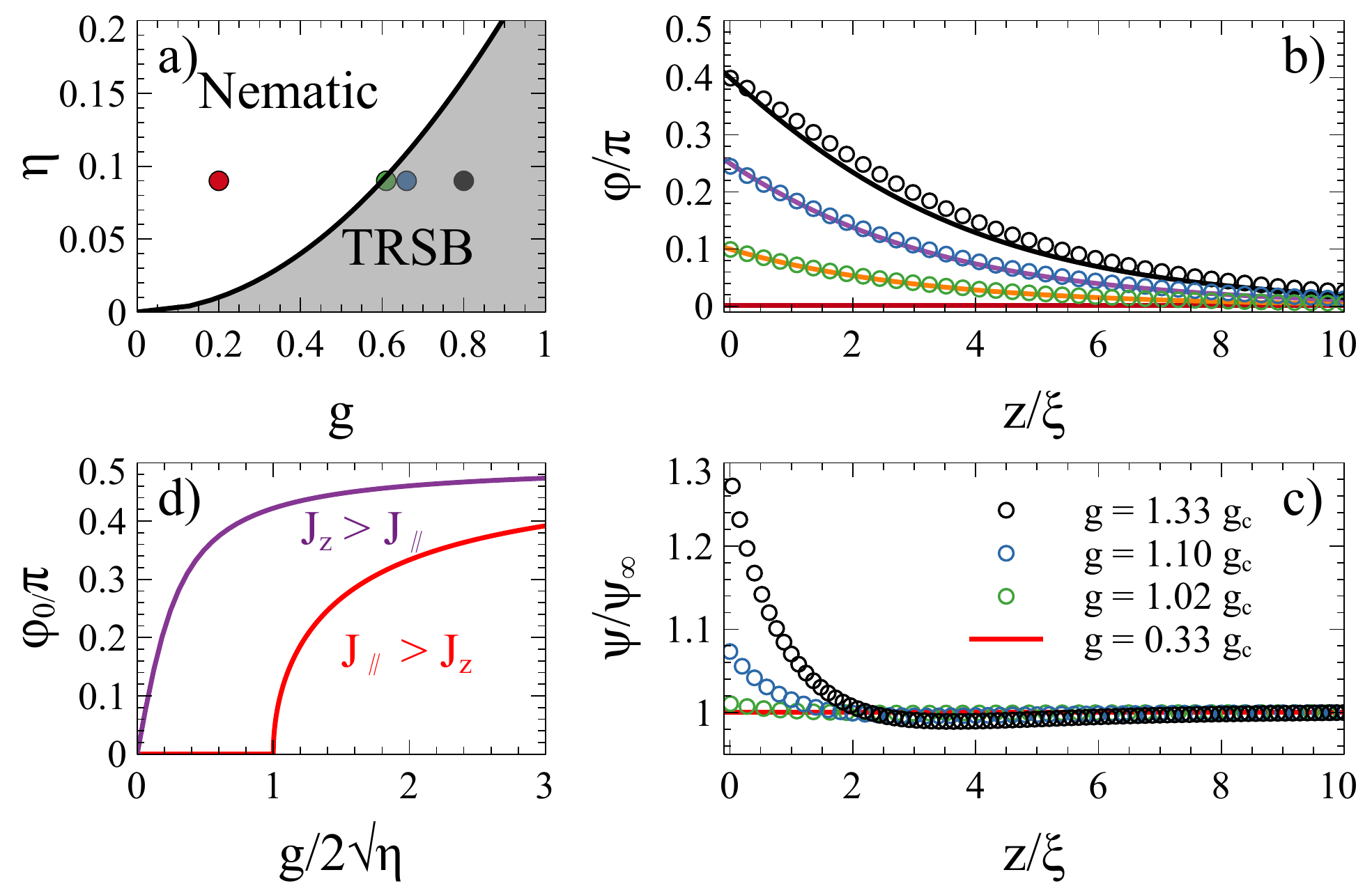}
\caption{a) Phase diagram for the onset of a surface TRSB phase. The separatrix $g_c=2\sqrt{\eta}$, marked in black, divides the diagram in a nematic phase for $g<g_c$ and a TRSB phase for $g>g_c$.  b) Phase $\varphi$ and c) amplitude $\psi$ versus the transverse direction $z$ for $\eta=0.09$: empty dots refer to the exact numerics and continuous lines in b) to Eq.~(\ref{Eq:soliton}). d) Surface phase $\varphi_0$ versus coupling for boundary conditions Eq.~(\ref{Eq:IntJparallel}) and Eq.~(\ref{Eq:IntZ}). 
\label{Fig:phaseD}}
\end{figure}

In case the magnetic order is in-plane, we have $g=\chi(\kappa J_z)^2n_{\rm imp}/(2|a|\psi_\infty^2n^2)$ and $U(\varphi)=\sin^2(\varphi)/4$, that provide the boundary condition $\varphi_0'=-g\sin(2\varphi_0)/4$. The solution for the phase reads 
\begin{equation}\label{Eq:soliton}
\varphi(x)=2{\rm arctan}\left[\tan(\varphi_0/2)e^{-\sqrt{\eta}x}\right],
\end{equation}
that represents a kink that matches the solution $\varphi_0$ at the origin with the asymptotic one $\varphi_\infty=0$. The boundary condition is solved by $\varphi_0={\rm arccos}(2\sqrt{\eta}/g)$ and  the associated free energy reads $\delta F=-g(1-2\sqrt{\eta}/g)/4$. A critical line $g_c=2\sqrt{\eta}$ separates a nematic solution $\varphi_0=0$ for $g<g_c$ and a TRSB solution $\varphi_0={\rm arccos}(2\sqrt{\eta}/g)$ for $g>g_c$, as shown in the phase diagram Fig.~\ref{Fig:phaseD}a). This way, for sufficiently strong coupling a surface TRSB state occurs with surface solution $\boldsymbol{\psi}_0\propto (1,e^{i\varphi_0})$. We numerically solve the coupled equations for amplitude and phase, and find an excellent agreement \cite{Note1}. The critical coupling  $g_c=2\sqrt{\eta}$ is matched exactly. The solution for the phase is shown in Fig.~\ref{Fig:phaseD}b) and closely matches Eq.~(\ref{Eq:soliton}), especially for small $\eta$. The amplitude is shown in Fig.~\ref{Fig:phaseD}c) and as expected varies on the scale $\xi$, whereas the phase varies on the scale $\xi/\sqrt{\eta}\gg\xi$. The purely chiral solution $\varphi_0=\pi/2$ is asymptotically reached for large $g$.  By inspection of Fig~\ref{Fig:phaseD}c) we also conclude that for a quasi 2D system satisfying $\xi>L$, $\boldsymbol{\psi}$ can be assumed constant and the results of Ref.~\cite{chirolli2017time-reversal} apply.  

\subsection{$J_z>J_\parallel$}

In case the magnetic order is out of plane (or in the non-interacting case) the boundary interaction is $U(\varphi)=\sin(\varphi)$ and $g=\kappa J_zn_{\rm imp}/(2n|a|\psi_\infty^2)$. It is clear that the out-of-plane magnetization favors a surface chiral OP and a TRSB solution always exists, as long as $g\neq 0$. The kink solution Eq.~(\ref{Eq:soliton}) applies and the value of the surface phase $\varphi_0$ is found by the boundary conditions $\varphi'_0=-2g\cos(\varphi_0)$, so that $\varphi_0={\rm arctan}(2g/\sqrt{\eta})$, that is nonzero for every $g>0$ and asymptotically reach the chiral solution $\varphi_0=\pi/2$. Furthermore, comparison to the $J_z<J_\parallel$ case shows that, for nominally equal coupling $g$, the chiral solution is obtained for much weaker coupling in the case $J_z>J_\parallel$ (see Fig.~\ref{Fig:phaseD}d)). 

\section{Discussion}

For Cr adatoms an almost isotropic spin exchange is predicted on Bi$_2$Se$_3$ \cite{chotorlishvili2014magnetic}.  Considering that magnetic adatoms tend to sit on precise microscopic lattice sites, either substitutional or interstitial, and that the nematic phase favors the crystallographic directions, the RKKY interaction cannot be completely ruled out on the basis of its peculiar directional dependence. In this case, an interacting picture applies and  a minimum density is required to trigger a surface TRSB solution if the order is in-plane. On the other hand, for magnetic adatoms characterized by $J_z\gg J_\parallel$, like Fe on Bi$_2$Te$_3$ \cite{eelbo2014strong}, a surface TRSB solution always arises and a relatively high impurity concentration can also be tolerated, owing to the predicted out-of-plane order. In this case, $g\simeq \mu J_z n_{\rm imp}/T_c^2$. Assuming $J_z/a^2\simeq 1~ {\rm meV}$, we find $g\simeq 10^{-2} n_{\rm imp}\xi^2$. It is important to stress that when magnetic impurities align no pair-breaking spin randomization takes place. In conclusion, we show how magnetic impurities on the surface of a nematic odd-parity superconductor can stabilize a surface chiral solution.

\section{Acknowledgments}

L.C. is thankful to E. Ercolessi for extensive and fruitful discussions, to T. Kvorning for a careful reading and relevant comments and he is grateful to F. Guinea for financial support through  funding from the European Commission under the Graphene Falgship, contract CNECTICT-604391. L.C. also acknowledges the European Commission for funding through the MCSA Global Fellowship grant TOPOCIRCUS-841894.

\bibliography{Bibfile}{}

\pagebreak


\onecolumngrid
\begin{center}
\textbf{\large Supplementary Material: Surface chiral superconductivity in odd-parity nematic superconductors with magnetic impurities
\\ [1cm]}
\end{center}
\twocolumngrid

\setcounter{equation}{0}
\setcounter{figure}{0}
\setcounter{table}{0}
\setcounter{page}{1}
\makeatletter
\renewcommand{\thesection}{S\arabic{section}}
\renewcommand{\theequation}{S\arabic{equation}}
\renewcommand{\thefigure}{S\arabic{figure}}
\renewcommand{\bibnumfmt}[1]{[S#1]}
\renewcommand{\citenumfont}[1]{S#1}
\setcounter{section}{0}

\section{Tight-binding model} 
\label{App:TB}

Here we provide details of the tight binding model used to calculate the band structure of the nematic superconductor. The Hamiltonian ${\cal H}_0({\bf k})$ can be seen as the expansion around the $\Gamma$ point at ${\bf k}=0$ of an extended tight-binding Hamiltonian. We assume the system to be constituted by two triangular lattices one on top of the other, labeled as $\sigma=T,B$, repeated on top of each other along the out-of-plane direction. Denoting as $c_{\sigma,s,i,i_z}$ the fermionic annihilation operators of layer $\sigma$, with spin $s$ at site $i$ in the triangular lattice and bilayer $i_z$, the real space tight-binding Hamiltonian of the bilayer structure reads
\begin{eqnarray}\label{Eq:TBmodel}
H&=&t_1\sum_{i,i_z,\sigma,\sigma',s}c^\dag_{\sigma,s,i,i_z}c_{\sigma',s,i,i_z}(\sigma_x)_{\sigma,\sigma'}\nonumber\\
&+&t_2\sum_{<ij>,i_z,\sigma,\sigma',s}c^\dag_{\sigma,s,i,i_z}c_{\sigma',s,j,i_z}(\sigma_x)_{\sigma,\sigma'}\nonumber\\
&+&i\lambda\sum_{<ij>,i_z,\sigma,s,s'}c^\dag_{\sigma,s,i,i_z}c_{\sigma,s',j,i_z}\sigma({\bf s}_{s,s'}\times{\bf d}_{ij})_z\nonumber\\
&+&t_z\sum_{i_z,i,\sigma,s}c^\dag_{\sigma,s,i,i_z}c_{\sigma',s,j,i_z+1}(\sigma_-)_{\sigma,\sigma'}+{\rm H.c.},
\end{eqnarray}
with $\sigma_i$ Pauli matrices in the layer space, $s_i$ Pauli matrices in the spin space, ${\bf d}_{ij}$ the unit vector pointing about the direction ${\bf r}_i-{\bf r}_j$. By Fourier transforming the tight-binding Hamiltonian and expanding around ${\bf k}=0$ we obtain a relation between the tight-binding coefficients and the parameters of the Hamiltonian Eq.~(1) in the main text, $m=t_1+6t_2$, $v=3\lambda a$, $v_z=a_zt_z$, with $a$ and $a_z$ the in-plane and out-of-plane lattice spacings. The parameters used in the numerical calculations of the bands are $t_1/t=-4.05$, $t_2/t=0.75$, $\lambda/t=0.5$, $t_z/t=-0.6$, $\mu/t=0.25$, $\Delta/t=0.03$, where $t$ is a generic hopping amplitude scale set to one. Additionally, we used $M_y/t=0.05$ with $M_y$ a Zeeman field along $\hat{y}$.

\section{Surface Majorana flat band}

The surface termination is implemented by the boundary condition $(1+\sigma_z)\phi(z=0)=0$ \cite{hsieh2012majorana}. The surface solutions are found by first solving the problem $H_{\rm BdG}({\bf k}=0,-i\partial_z)\phi=0$. Assuming $\psi_x=\Delta$, $\psi_y=0$, a Kramers pair of Majorana fermions is found with wavefunction
\begin{equation}\label{Eq:Mwf}
\phi_\alpha(z)=e^{-z/\xi}\left(\begin{array}{c}
\sin(k_Fz)\\
\sin(k_Fz+\gamma)
\end{array}\right)_\sigma|\phi_\alpha\rangle,
\end{equation}
with $k_F=\sqrt{\mu^2-m^2}/v$, $e^{i\gamma}=m/\mu+i\sqrt{1-m^2/\mu^2}$, $\xi=v/\Delta$, and $|\phi_\alpha\rangle$ eigenstates of $s_y\tau_y$ with eigenvalue -1, and $\alpha=\pm 1$, $|\phi_\alpha\rangle=\frac{1}{2}\left(\begin{array}{c}1\\ i \alpha\end{array}\right)_s\otimes \left(\begin{array}{c}1\\ -i \alpha\end{array}\right)_\tau$. The surface Hamiltonian is then obtained by taking the matrix elements of the term $v\sigma_z(k_zs_y-k_ys_x)\tau_z$ onto the Majorana states $\phi_\alpha$. The Green's function of the Majorana flat band is written as
\begin{equation}
G(x,y)=\frac{\sin(k_Fx)}{i\pi^2 x}\int_0^{k_Fy}dz\frac{\cos(z)y\omega_n/v_M+z\sin(z)\hat{\alpha}_y}{(y\omega_n/v_M)^2+z^2}.
\end{equation}

\subsection{Nodes contribution}

Here we consider the surface contribution arising from bulk nodal states. We introduce the manifestly covariant Bloch basis (MCBB) $|\psi_{\rm cb}^1\rangle_{\bf k}$, $|\psi_{\rm cb}^2\rangle_{\bf k}$ describing the conduction band. These states are chosen to be fully spin polarized along the $z$ direction at the origin in ${\bf k}$ space \cite{fu2015parity-breaking,venderbos2016odd-parity}. Rescaling the momentum as $k_z\to k_zv/v_z$, the gap matrix projection on the conduction band takes the form $\hat{\Delta}=\frac{v}{\mu}(\Delta_x F_x({\bf k})+\Delta_y F_y({\bf k}))$, with 
\begin{equation}
F_x=k_y\tilde{s}_z-k_z\tilde{s}_y,\qquad F_y=-k_x\tilde{s}_z+k_z\tilde{s}_x,
\end{equation}
where $\tilde{s}_i$ are Pauli matrices spanning the subspace $\{|\psi_{{\rm cb}}^1\rangle_{{\bf k}}, |\psi_{{\rm cb}}^2\rangle_{{\bf k}}\}$. Assuming the nematic director to be oriented along the $x$ direction, the MCBB around at the nodes ${\bf k}_\pm=(\pm k_F,0,0)$ in the basis $\{c_{T\uparrow},c_{B\uparrow},c_{T\downarrow},c_{B\downarrow}\}$  reads
\begin{equation}
|\psi_{\rm cb}^1\rangle_\pm=\frac{1}{2}\left(\begin{array}{c}
\sqrt{1+\eta}\\
\sqrt{1+\eta}\\
\pm i\sqrt{1-\eta}\\
\mp i\sqrt{1-\eta}
\end{array}\right),
|\psi_{\rm cb}^2\rangle_\pm =\frac{1}{2}\left(\begin{array}{c}
\mp i\sqrt{1-\eta}\\
\pm i\sqrt{1-\eta}\\
\sqrt{1+\eta}\\
\sqrt{1+\eta}
\end{array}\right),
\end{equation}
where $\eta=m/\mu$. 

\begin{figure}[t]
\includegraphics[width=0.45\textwidth]{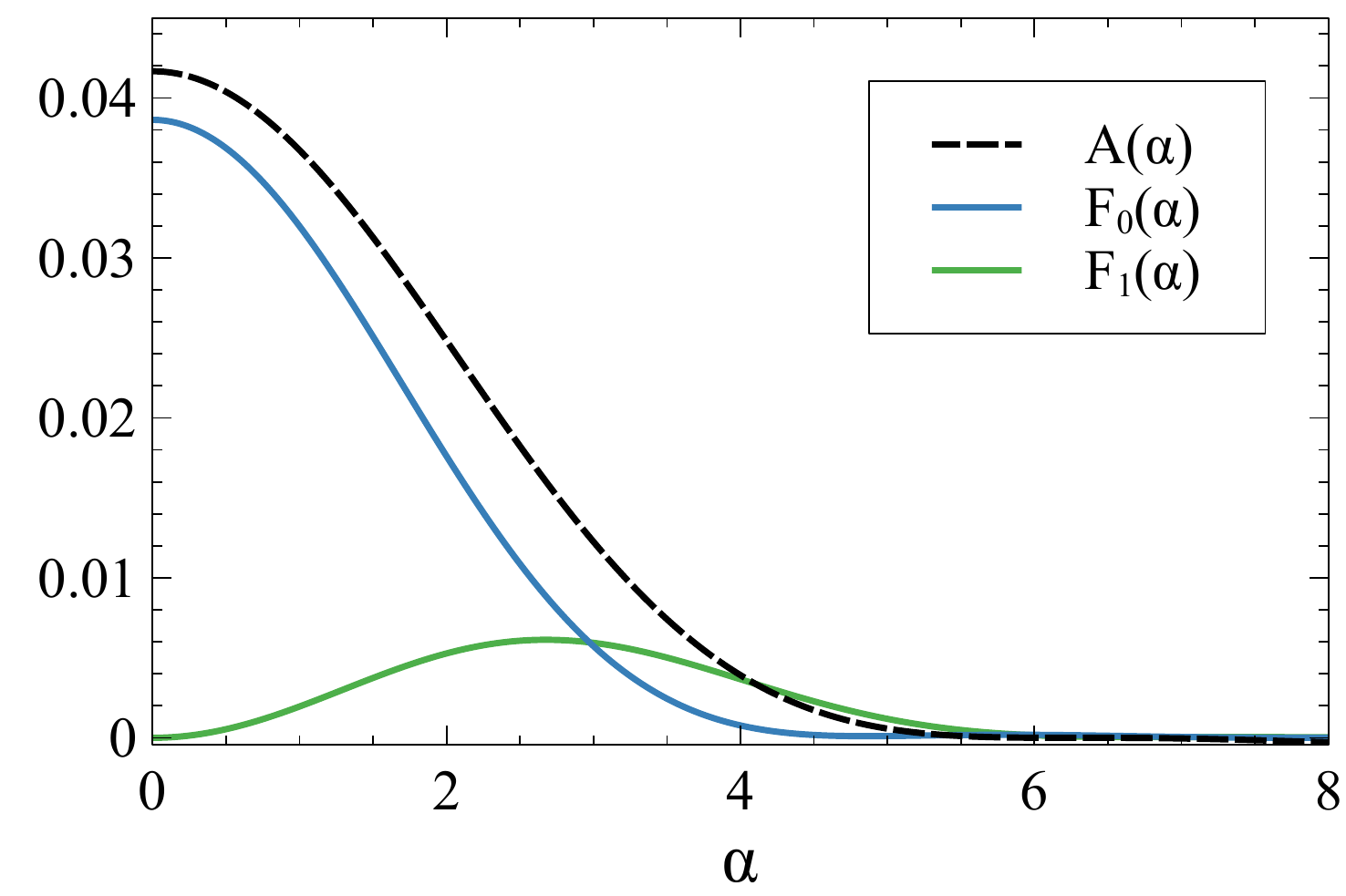}
\caption{The function $F_0$, $F_1$, and $A$ that characterize the RKKY dependence on the rescaled impurity distance. \label{Fig4}}
\end{figure}

We then project the BdG Hamiltonian Eq.~(2) onto the MCBB and expand it around the nodal points ${\bf k}_\pm=(\pm k_F,0,0)$, keeping the MCBB constant, and find
\begin{equation}\label{Eq:H4x4}
h^\pm_{\bf k}=\boldsymbol{\epsilon}_{\bf k}\cdot{\bf P}^\pm,\qquad
{\bf P}^\pm=(\pm \tau_z,\tau_x\tilde{s}_z,\tau_x\tilde{s}_y)
\end{equation}
with $\epsilon_i=\tilde{v}_ik_i$,  $\tilde{v}_x=k_Fv^2/\mu$, $\tilde{v}_y=\Delta v/\mu$, $\tilde{v}_z=v_z\Delta/\mu$. The Green's function takes the form
\begin{equation}
G({\bf r})=\sum_{\pm}e^{\pm ik_Fx}\mathbb{P}^\pm_{ij}\sum_{\bf k}e^{i{\bf k}\cdot{\bf r}}[(i\omega_n-h^\pm_{\bf k})^{-1}]_{ij},
\end{equation}
where $\mathbb{P}^\pm_{ij}=|\psi_{\rm cb}^{i}\rangle_{\pm}\langle\psi^j_{\rm cb}|$. The natural energy cutoff is set by the bulk gap $\Delta$. Introducing the dimensionless position $\boldsymbol{\rho}=\mu(x/(k_F\xi),y,zv/v_z)/v$, the Green's function can be written as
\begin{eqnarray}
G(\boldsymbol{\rho})&=&\frac{\nu_F\mu^2}{i(k_Fv)^2}\sum_{\pm}e^{\pm ik_Fx}\left[f_0\left(\rho,\frac{\omega_n}{\Delta}\right)\right.\nonumber\\
&+&\left.f_1\left(\rho,\frac{\omega_n}{\Delta}\right)\tilde{{\bf r}}\cdot{\bf P}^\pm\right]_{ij}\mathbb{P}^\pm_{ij},
\end{eqnarray}
where $\tilde{\bf r}=(x \tilde{v}_x/\tilde{v}_z,y\tilde{v}_y/\tilde{v}_z,z)/r$,   $\nu_F=\mu k_F/(2\pi^2 v v_z)$ is the density of states at the Fermi level of the bulk Hamiltonian and 
\begin{eqnarray}
f_0(\alpha,\beta)&=&\frac{\beta}{\alpha}\int_0^{\alpha}\frac{d\zeta \zeta^2j_0(\zeta)}{\zeta^2+\alpha^2\beta^2},\\
f_1(\alpha,\beta)&=&\frac{1}{\alpha^2}\int_0^{\alpha}\frac{d\zeta \zeta^3j_1(\zeta)}{\zeta^2+\alpha^2\beta^2},
\end{eqnarray}
with $j_n(\zeta)$ the spherical Bessel function of the first kind. The integrals can be done analytically and are expressed in terms of sine and cosine integrals and hyperbolic functions. At $T=0$ the sum over frequency can be transformed to an integral, $T\sum_n\to \int \frac{d\omega}{2\pi}$. The full expressions are given in the main text in terms  of the functions
\begin{eqnarray}
A_i(\rho)&=&F_0(\rho)-\tilde{r}_i^2F_1(\rho), \qquad i=0,x,y,\\
A_z(\rho)&=&F_0(\rho)+\tilde{r}_z^2F_1(\rho), 
\end{eqnarray}
where $\tilde{r}^2_i=\tilde{\bf r} . V_i . \tilde{\bf r}$, $V_0=1$, $V_x={\rm diag}(-1,1,1)$, $V_y={\rm diag}(1,-1,1)$, $V_z={\rm diag}(1,1,-1)$ and
\begin{eqnarray}
F_i(\alpha)&=&\int_{-\infty}^\infty\frac{d\beta}{2\pi}f^2_i(\alpha,\beta).
\end{eqnarray}
The integral over the frequency is best performed in the complex plane and we find
\begin{eqnarray}
F_0(\alpha)&=&\frac{1}{\alpha^5}\int_0^\alpha\frac{dz_1dz_2 z_1z_2}{2(z_1+z_2)}\sin(z_1)\sin(z_2),\\
F_1(\alpha)&=&\frac{1}{\alpha^5}\int_0^\alpha\frac{dz_1dz_2 z^2_1z^2_2}{2(z_1+z_2)}j_1(z_1)j_1(z_2).
\end{eqnarray} 
They both admit an analytical solution in terms of sinusoidals and sine and cosine integral functions. $F_0$ can be approximated as
\begin{eqnarray}
F_0(\alpha)\lesssim A(\alpha) =\frac{\sin^3(\alpha/2)}{3\alpha^3}.
\end{eqnarray}
The function $F_0\gg F_1$ for $\alpha<1$ and they are both suppressed for larger argument, as it is shown in Fig.~\ref{Fig4}.  

\section{Numerical solution of the GL equations for a semi-infinite system}
\label{App:GLequations}

Here we discuss details of the GL equations in the $z>0$ half-plane. In general the two-component order parameter is specified by four real functions, two amplitudes and two phases. In absence of external magnetic fields the global phase can be gauged away. The relative amplitude fix the direction in the space of the nematic director and we assume an asymptotic solution $\boldsymbol{\psi}_{\infty}=\psi_\infty(1,1)/\sqrt{2}$. We are then left with a global amplitude and a relative phase, $\boldsymbol{\psi}=\psi(e^{-i\varphi/2},e^{i\varphi/2})/\sqrt{2}$. The GL free energy then reads
\begin{eqnarray}
F&=&\int_0^Ldz\left[\beta_z(\partial_z\psi)^2+\frac{\beta_z}{4}\psi^2(\partial_z\varphi)^2\right]-|a|\psi^2+b\psi^4\nonumber\\
&+&b'\psi^4\sin^2(\varphi)-\frac{\chi n_{\rm imp}\kappa^2 J_\parallel^2}{n^2}\psi_0^4\sin^2(\varphi_0),
\end{eqnarray}
with $S$ that surface area. We rescale the field as $\psi=\psi_\infty f$, with the asymptotic amplitude $\psi_\infty^2=|a|/(2b)$, and the position $z=\xi x$ by the coherence length $\xi^2=\beta_z/(2|a|)$. Subtracting the asymptotic bulk free energy $F_0=-|a|\psi_\infty^2/2$ we are left with
\begin{equation}
\delta F=\frac{F-F_0}{2\psi_\infty^2|a|}\int_0^\infty dx{\cal F}(x,f,f',\varphi,\varphi')
\end{equation}
with 
\begin{eqnarray}
{\cal F}&=&(f')^2+\frac{1}{4}(1-f^2)^2+\frac{f^2}{4}((\varphi')^2+\eta f^2\sin^2(\varphi))\nonumber\\
&-&\frac{g}{4}f_0^4\sin^2(\varphi_0)\delta(x)
\end{eqnarray}
where $g=\chi n_{\rm imp}(\kappa J_\parallel/n)^2\psi_\infty^2/(2|a|)$ and we took the boundary condition inside the integral as a delta function by slightly extending the integral to negative $x$ values.

At infinity the solution are $f=1$ and $\varphi=0,\pi$. For $g=0$ the GL equations for $f$ and $\varphi$ are obtained by extremizing the free energy and read 
\begin{equation}\label{AppGLeqs}
\frac{\partial {\cal F}}{\partial f}-\frac{d}{dx}\frac{\partial {\cal F}}{\partial f'}=0,\qquad \frac{\partial {\cal F}}{\partial \varphi}-\frac{d}{dx}\frac{\partial {\cal F}}{\partial \varphi'}=0.
\end{equation}
They are explicitly given by
\begin{eqnarray}
f''&=&\frac{f}{4}(\varphi')^2-\frac{f}{2}(1-f^2)+\frac{\eta}{2} f^3\sin^2(\varphi),\label{App:GLamp}\\
f^2\varphi''&=&-2ff'\varphi'+\frac{\eta}{2}f^4\sin(2\varphi),\label{App:GLphase}
\end{eqnarray}
By requiring the variations $\delta f$, $\delta \varphi$ to be zero only at infinity, $\delta f(\infty)=\delta \varphi(\infty)=\delta \Theta(\infty)=0$ we obtain the additional constraints
\begin{equation}
\left.\frac{\partial {\cal F}}{\partial f'}\right|_0=-g \left.\frac{\partial U}{\partial f}\right|_0,\qquad \left.\frac{\partial {\cal F}}{\partial \varphi'}\right|_0=-g \left.\frac{\partial U}{\partial \varphi}\right|_0.
\end{equation}
It follows that the contact interaction proportional to $g$ generates the boundary conditions
\begin{eqnarray}
\varphi'_{0}&=&-\frac{g}{4}f_0^2\sin(2\varphi_0),\label{BCph}\\
f'_{0}&=&-\frac{g}{4}f_0^3\sin^2(\varphi_0)\label{BCamp}.
\end{eqnarray}
An exact analytical solution of Eqs.~(\ref{App:GLamp},\ref{App:GLphase}) is unfortunately not available.

\begin{figure}[t]
\includegraphics[width=0.45\textwidth]{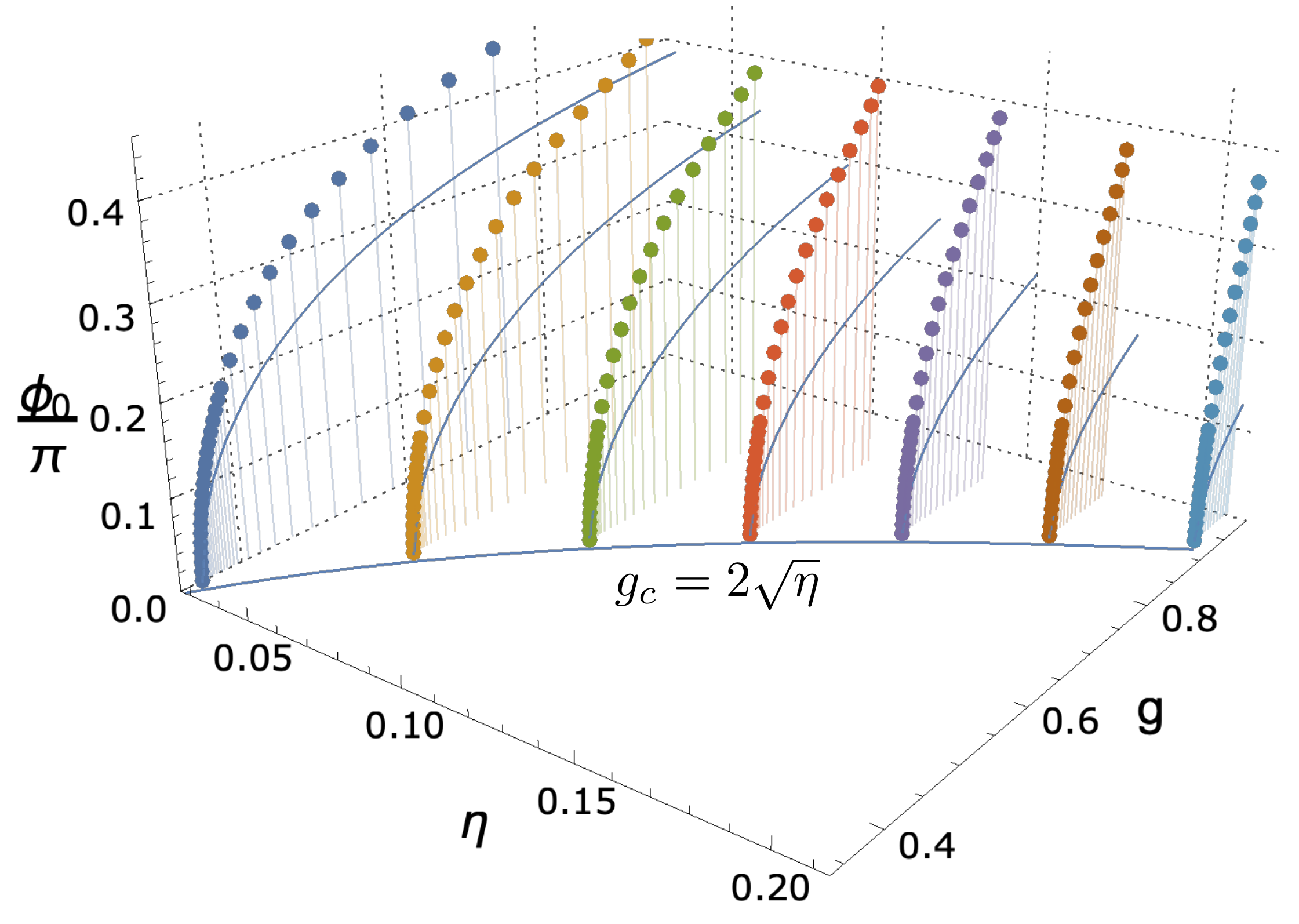}
\caption{Phase diagram for the onset of a surface TRSB phase for $g>g_c$. The separatrix $g_c=2\sqrt{\eta}$ is plotted in the $(g,\eta)$ plane. The surface value of the phase $\varphi_0$ is obtained by numerical solutions of the coupled equations (\ref{App:GLamp},\ref{App:GLphase}) with the boundary conditions Eqs.~(\ref{BCph},\ref{BCamp}). \label{Fig:PD}}
\end{figure}

We then proceed to numerically solve Eqs.~(\ref{App:GLamp},\ref{App:GLphase}) for $\eta$ sufficiently small. We first generate numerical solutions fixing the boundary conditions far away from $x=0$. We then find the curves satisfying $\varphi_0'f_0\tan(\varphi_0)=f_0'$, and extract the relative value of $g$. In Fig. 3b) and 3c) of the main text we show solutions for the amplitude and phase for a given value $\eta=0.09$ and three values of $g$, that closely match the kink solution for the phase. We then proceed to extract the values of $\varphi_0$ as a function of $g$. The result is shown in Fig.~\ref{Fig:PD}. The basal line in the plane $(\eta,g)$ is given by the separatrix $g_c=2\sqrt{\eta}$ and shows that the condition for a TRSB phase $g>2\sqrt{\eta}$ is exactly matched. Numerical solutions are shown by full dots and are compared with the corresponding approximate analytical solutions $\varphi_0={\rm arccos}(2\sqrt{\eta}/g)$, shown as full lines. We see that the analytical formula underestimates the values of $\varphi_0$. We then conclude that a surface TRSB phase can be stabilized by magnetic impurities for sufficiently strong coupling.

\end{document}